# Effect of bed temperature on solute segregation and mechanical properties in Ti-6Al-4V produced by selective laser melting

S. Pedrazzini[1*], M. E. Pek[1*], A. K. Ackerman[1], Q. Cheng[1], H. Ali[2,3], H. Ghadbeigi[2], K. Mumtaz[2], T. Dessolier[1], T.B. Britton[1], P. Bajaj[4], E. Jägle[4], B. Gault[1,4], A. J. London[5], E. Galindo-Nava[6]

[1] Department of Materials, Imperial College London, South Kensington Campus, Exhibition Road, SW7 2AG, London, UK.

[2] Department of Mechanical Engineering, University of Sheffield, Western Bank, Sheffield, S1 3JD, UK.

[3] University of Engineering & Technology, Jamrud Road, Peshwar Khyber Pakhtunkhwa, Pakistan.

[4]Max-Planck Institute für Eisenforschung, Max-Planck Straße 1, 40237, Düsseldorf, Germany.

[5] UK Atomic Energy Authority, Culham Science Centre Abingdon OX14 3EB, UK.

[6] Department of Materials Science and Metallurgy, University of Cambridge, 27 Charles Babbage Road, CB3 0FS, Cambridge, UK.

*These two authors contributed equally.

**Abstract**

1. Introduction

Laser Powder Bed Fusion (LPBF) is an additive manufacturing (AM) process where a high-intensity laser melts and fuses selected regions of powders deposited in cross-sectional layers. The near net-shape production of components minimises the need for specialised tooling and allows the manufacture of increasingly complex designs, better tailored to the requirements of specific applications. These characteristics are very attractive for manufacturing high-performance alloys in the aerospace industry. Titanium alloys, particularly Ti-6Al-4V, are used widely in the aerospace industry due to their high specific strength and hot corrosion resistance [1]. However, titanium alloys are expensive and time-consuming to produce using conventional cast and wrought methods [2]. LPBF is a viable alternative that allows the production of near-net shape bespoke components, reducing the cost and timeliness of substantial machining. However, several challenges have prevented this technology from reaching optimal commercial implementation. For instance, Ti-6Al-4V produced by LPBF, without further treatments such as shot peening or hot isostatic pressing (HIP), reportedly showed reduced ductility and fatigue resistance compared with its counterpart produced by conventional methods [3]. This is due to the large temperature gradients and fast solidification rates which generates high residual stresses [4] and, more notably, the formation of highly textured, metastable martensitic structures [5].

Common strategies to improve the mechanical properties of LPBF Ti-6Al-4V relies on post-production heat treatments designed to decompose the martensitic structure [5]. Vilaro *et al.* [6] studied the microstructure produced by quenching from the β phase field and subsequent heat treatments at, above, and below the β transformation temperature (transus). The highest ductility values were obtained when tempering close to the martensite start temperature (Ms) and the resulting microstructure was a mixture of α'+β+α with lamellar morphologies. Other authors have found analogous microstructures and mechanical properties using similar heat treatments [7–9]. In addition, the use of multiple lasers in LPBF produced similar microstructures and mechanical properties [10]. Additional processes to reduce the residual stresses in LPBF produced materials include methods such as electropolishing [11] and hot isostatic pressing (HIP), which can also be used to reduce residual porosity [12]. These processes typically occur after an initial heat treatment [13]. In recent work, a heated base-plate was used during LPBF production of Ti-64 to reduce residual stresses and decompose the martensitic structure [14]. Ali *et al.* found an increase in ductility with increasing temperature between 100-570 ˚C, then a subsequent decrease up to 770 ˚C.

Solute segregation in titanium alloys, particularly at interfaces and defects, has not been widely studied. The subject of segregation at interfaces has been discussed at length by Raabe *et al.* [15], in which the authors discuss how segregation at interfaces and defects may be used beneficially to encourage the nucleation of additional phases. Ackerman *et al.* [16] have previously observed molybdenum peaks at grain boundaries in Ti-6Al-2Zr-4Sn-6Mo, and regularly space concentrations of Zr, thought to be segregation to defects [17]. In materials other than titanium, such as aluminium, such segregation has been observed. In nanocrystalline Al, it has been suggested that oxygen pins the grain boundaries, limiting grain boundary movement [18].

The present work employs advanced characterisation techniques, including SEM, TEM and APT, to correlate the changes in microstructures with elevated substrate temperature. These results are compared to a nucleation model to try and explain the phenomena observed. It is found that complex phase transition sequences and solute redistribution behaviour control the final microstructure, and these are not comparable to standard LPBF techniques. Based on these observations, the mechanisms controlling the strength and ductility at different substrate temperatures are elucidated.

## 2. Experimental Methods

Gas-atomised powder of nominal chemical composition Ti-6Al-4V was provided by TLS Technik Spezialpulver and sieved to 15-45 µm in diameter. The powders were used to manufacture 6 blocks, 30 x 30 x 10 mm in size, through LPBF on a pre-heated substrate at 100, 370, 470, 570, 670 and 770 °C. This was completed using a Renishaw SLM125 system with a modified pre-heating platform. The build was completed using the default Renishaw SLM parameters: laser focus offset 0, hatch spacing 0.08 mm, contour spacing 0.2, layer thickness 50 µm, scanning strategy 90˚ alternative. The laser power and exposure time was varied between 120-200 W and 60-180 µs respectively. Further details of production, residual stress measurements and porosity measurements have been published as part of a previous manuscript [14].

### 2.1 Metallographic preparation and imaging

Samples were sectioned for metallographic examination vertically along the x-z axis of the printed specimens using a diamond saw, then hot-mounted in conductive bakelite (with carbon filler), polished initially with SiC grinding paper, then with 3 and 1 µm diamond paste and finally 0.04 µm colloidal silica to obtain a smooth surface finish. Scanning electron microscopy (SEM) was performed on sections of the samples using a Zeiss GeminiSEM 300 microscope. A working distance of 8.5 mm was used, operating at a beam current of 3 nA and a voltage of 5 kV. A combination of secondary and backscattered electron imaging was used, in order to fully exploit the surface sensitivity and Z-contrast. Electron backscatter diffraction (EBSD) measurements were performed with a Bruker eFlash$^{HD}$ EBSD camera inside a FEI Quanta 650 scanning electron microscopy (SEM) using a beam acceleration voltage of 20 kV and a probe current of ~10 nA. Each sample was analysed with an EBSP resolution of 320x240 pixels, samples at 570 and 770 °C were analysed with a step size of 40 nm and an EBSD camera exposure of 25 ms while the sample at 100 °C was captured with a step size of 200 nm and an EBSD camera exposure of 70 ms.

### 2.2 Transmission electron microscopy (TEM)

Transmission electron microscopy (TEM) was performed on 3 mm disc samples, which were cut using a diamond saw, then electropolished using a Struers Tenupol with a solution of 15% (by volume) perchloric acid in methanol, cooled with $LN_2$. Each specimen was examined using a Technai Osiris microscope with a Bruker Energy Dispersive X-ray detector (EDX). From the micrographs acquired, particle size distributions were determined through manual measurements. The matrix compositions of each specimen were measured by TEM-EDX, and any values presented here were obtained by taking the mean value over at least 10 measurements. Semi-quantitative data was obtained from EDX maps using the Cliff-Lorimer method.

### 2.3 Atom probe tomography (APT)

Atom probe tomography (APT) was performed on the samples using a LEAP 5000 XS in laser-pulsing mode. Specimens were prepared by in-situ lift out using a FEI Helios NanoLAB 600i dual-beam SEM-FIB, equipped with an Omniprobe micromanipulator. The Ga beam was used to prepare cantilevers, which were then welded onto the micromanipulator using Ga-beam deposited Pt, extracted and mounted onto Cameca Silicon flat-top coupons. Specimens were then sharpened using the Ga-beam until they were below 100 nm in diameter, then cleaned using 5 kV Gallium, to remove the region minimise the ion beam damage. Analysis conditions were varied based on the specimen's profile within a selected range: temperatures between 40–60 K, laser energy in the range of 10–30 pJ, and a pulse frequency between 100 and 200 kHz. Reconstructions were performed using the Cameca Integrated Visualisation and Analysis Software (IVAS 3.8.4).

## 3. Results

### 3.1 Mechanical characterisation

Mechanical test results were published as part of a previous study [14] discussing the methods for alleviating internal stresses caused by additive layer manufacturing through heating the sample bed. **Figure 1 (a)** summarises the values of ductility and how they are influenced by the substrate temperature. When the substrate is heated to 100°C, the elongation is ~6%, which increases and peaks at ~10% at 570°C, then sharply decreases to 0% (brittle failure, no ductility) at 770°C. **Figure 1 (b)** shows the change in ultimate tensile stress (UTS) with respect to substrate temperature. It can be seen that above 670°C there is a sharp decrease in the UTS of the material, which can be related to the similar loss in ductility, due to brittle failure. **Figure 1(c)** shows the full tensile curves of the material. Whilst sample printed at a bed temperature of 100°C and 570°C have similar curves, with minimal elongation at 100°C, samples tested from a bed temperature of 770°C are incredibly brittle.

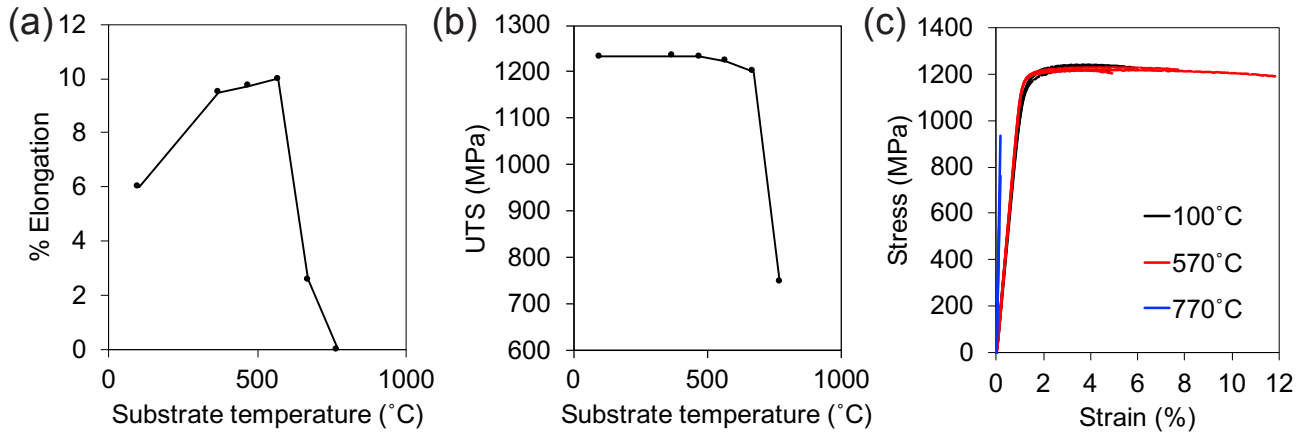

*Figure 1: (a) % elongation vs substrate temperature during LPBF processing, measured during room temperature tensile tests, performed at $10^{-4}$ $s^{-1}$ strain rate. The sample made by pre-heating the powder bed at 770 °C was brittle. (b) ultimate tensile strength (UTS) vs substrate temperature. (c) tensile curves for samples printed at a bed temperature of 100 °C (black), 570 °C (red) and 770 °C (blue).*

### 3.2 Microstructural characterisation

EBSD was performed on the samples to better understand the mechanical testing observations. When building a sample on a heated substrate at 100°C, the microstructure consists primarily of α and α' phases (as α and α' have too similar a lattice parameter and crystal structure, EBSD could not be used to differentiate between them). Small pockets of β phase could be identified. EBSD also shows extensive presence of microtwins in the 100°C sample. In comparison, the sample produced on a heated substrate at 570°C shows no alterations in grain size β phase fraction, though there is a complete absence of microtwins. At 770°C no twins are visible and the α grain size is increased.

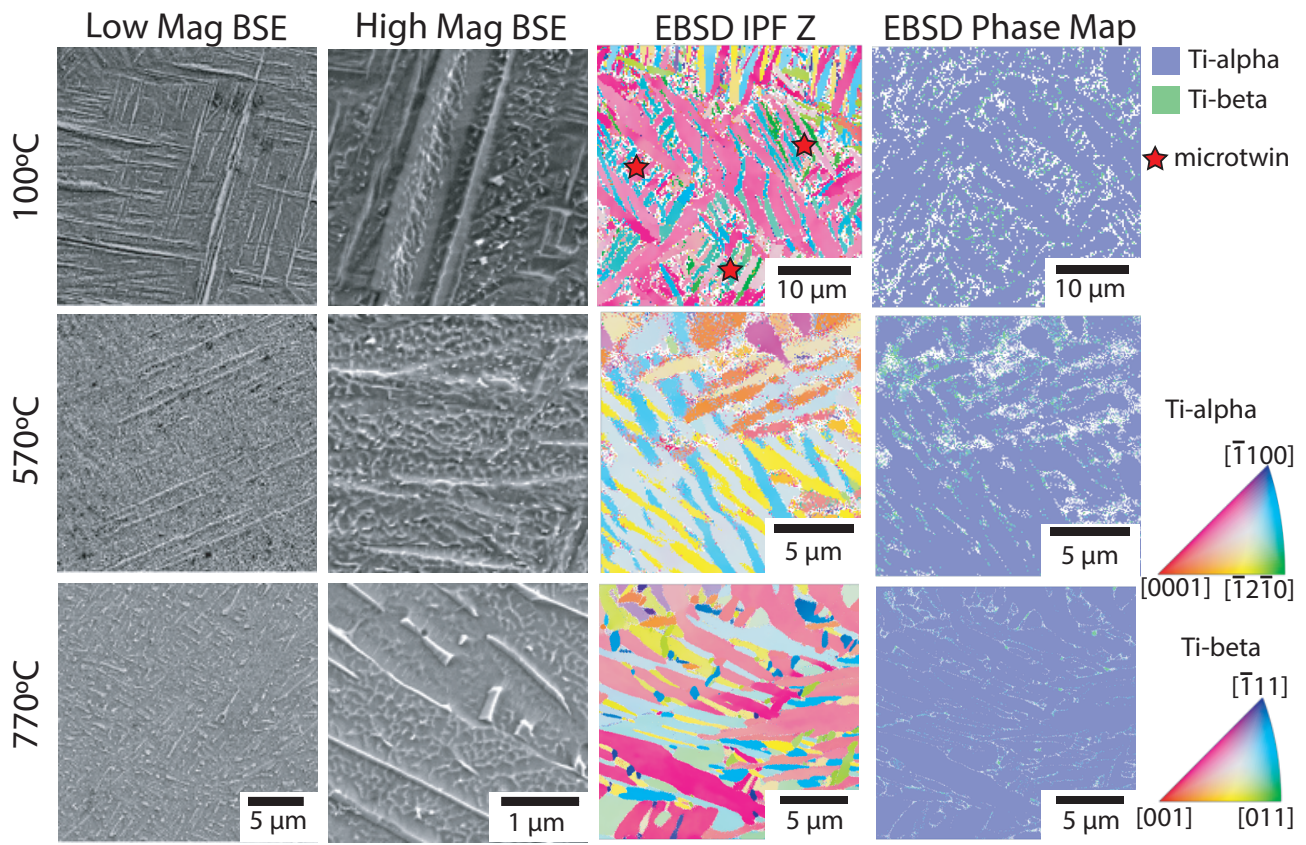

*Figure 2*: Backscattered electron micrographs showing the overall microstructure of the LBPF printed samples and details of the phases present at higher magnification. Micrographs were taken from a central region of the x-z axis (vertical sections). EBSD inverse pole figure (IPF) maps along the x-axis and a phase fraction map for each sample.

**3.2 Compositional analysis and fine scale microstructure**

Bright field TEM and EDX was performed on the samples and the results are shown in **Figures 3 to 5**. **Figure 3** are micrographs and EDX maps of the sample produced at a substrate temperature of 100°C. Microtwins are visible, which is consistent with the results shown in **Figure 2**. EDX maps could not resolve a possible segregation of solutes at the microtwin boundaries.

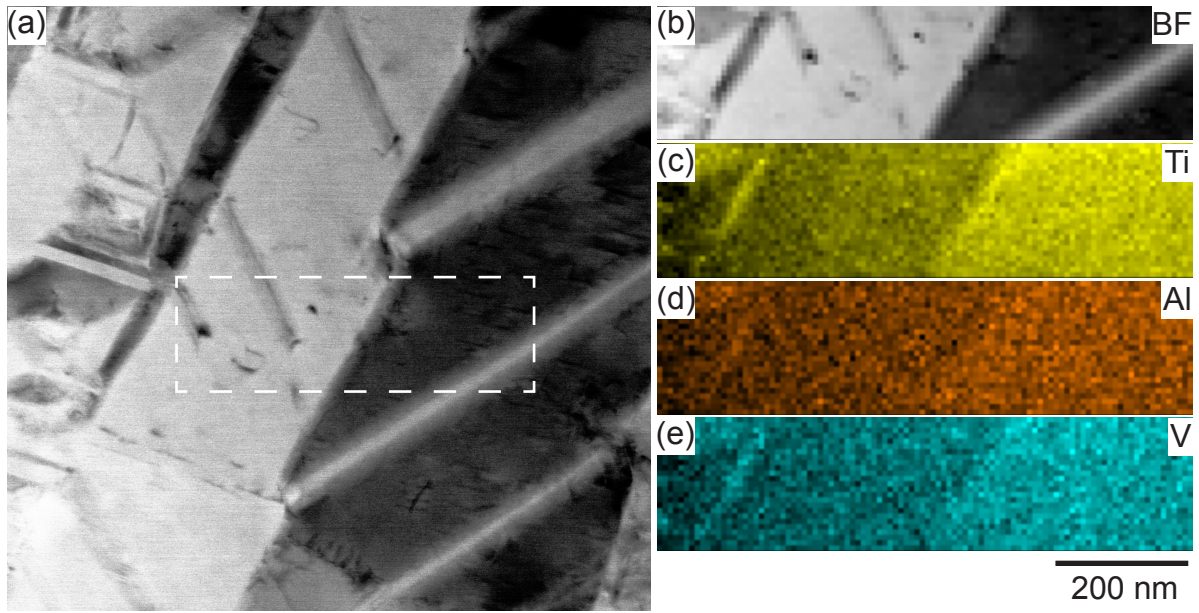

*Figure 3*: STEM-EDX results of the sample produced at 100°C. The foil is taken with the build axis out of the page. (a) STEM –BF of 100˚C sample with the dotted rectangle indicating the sampled EDX area; (b) Magnified STEM-BF; (c) Ti-K$\alpha$; (d) Al-K$\alpha$; (e) V-K$\alpha$

The sample produced at a substrate temperature of 570°C showed no microtwin boundaries (**Figure 4**) but EDX maps showed small quantities of nano-scale β phase (shown by the rejection of V from the α phase in the EDX) which were too small to be identified by EBSD in **Figure 2**. There is a presence of dislocations which is not seen in the 100 °C specimen, which are more substantial within the α lamellae.

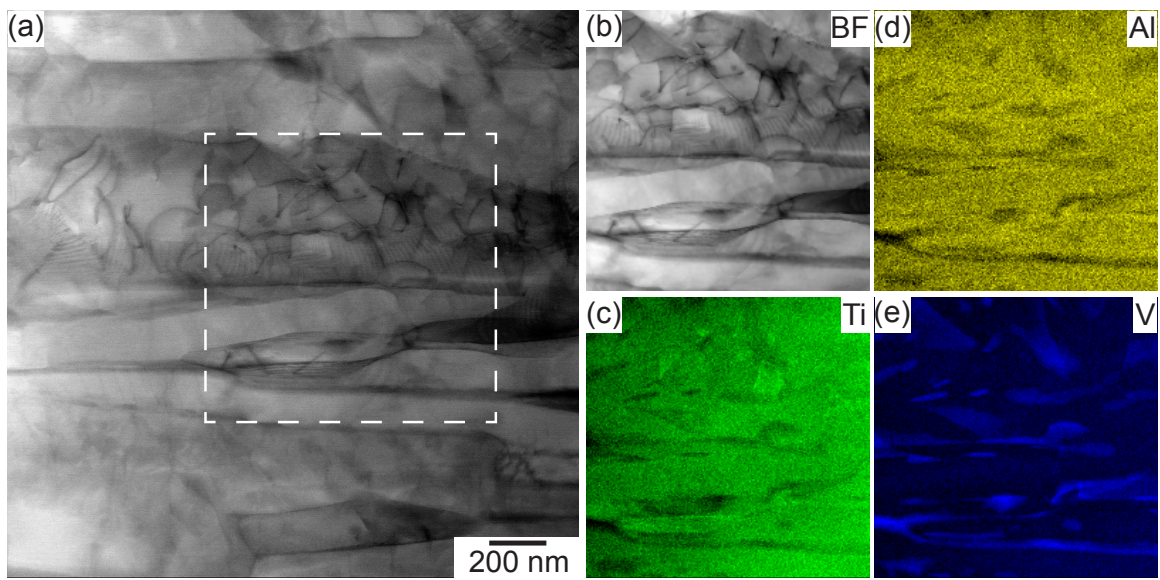

*Figure 4*: STEM-EDX results of the sample produced at 570°C. The foil is taken with the build axis out of the page. (a) STEM –BF of 570˚C sample with the dotted rectangle indicating the sampled EDX area; (b) Magnified STEM-BF; (c) Ti-K$\alpha$; (d) Al-K$\alpha$; (e) V-K$\alpha$

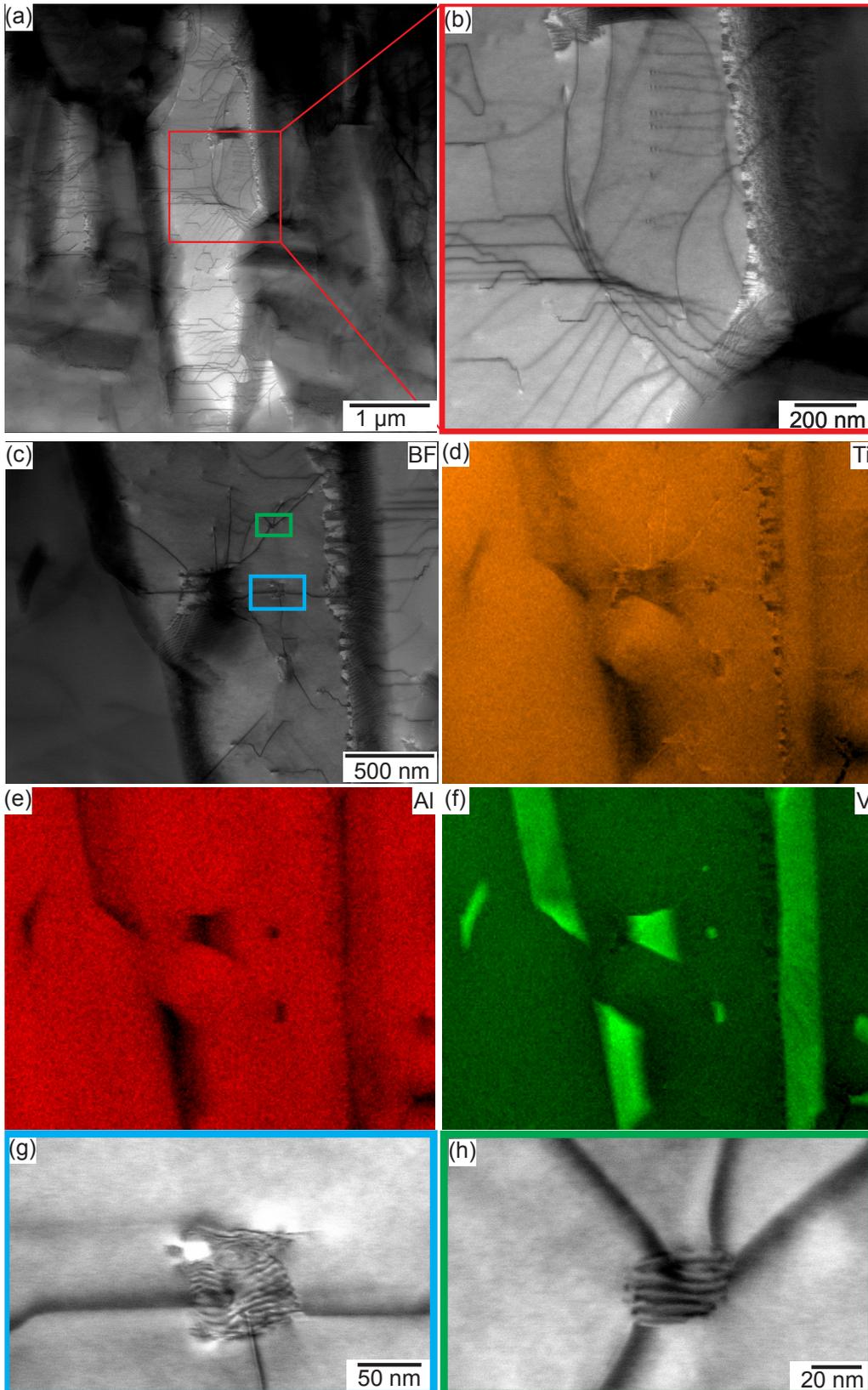

*Figure 5*: STEM-EDX results of the sample produced at 770°C. (a) STEM–BF of 770°C sample at low magnification to show β grain; (b) Magnified STEM-BF to show dislocation along preferential planes; (c) STEM–BF with (d) Ti-K$\alpha$; (e) Al-K$\alpha$; (f) V-K$\alpha$ (g-h) magnified view of two $\beta$-Ti grains highlighted in (c). The foil is taken with the build axis out of the page.

The sample produced at a substrate temperature of 770°C also had nano-scale faceted β grains (**Figure 5**). **Figure 5(a)** and **(b)** also showed networks of dislocations, tangled and evidently cross-slipped along specific crystal orientations, which is visible from the 60° angles between them (in this observed axis). **Figures 5 (c)** and **(g-h)** show the nano-scale $\beta$ phase. **Figures 5 (d)** to **(f)** also shows that the dislocations have segregation of V and that V-rich phases, such as nano-β, could be pinning dislocations. There is also evidence of Ti segregation along the dislocation lines, **Figure 5 (d)**, which could again be effecting dislocation mobility.

### 3.3 Fine scale phase segregation

In order to investigate the fine scale compositional variations, APT was used. **Figure 6** shows an atom probe reconstruction from each sample. The local concentration of V and Al is shown as the iso-concentration surfaces of 4 at% and 12 at% respectively, to allow a distinction between the $\alpha'$-Ti and $\alpha$-Ti phases, which have the same crystal structure but differ in chemical composition and lattice parameter [19]. The change in the relative amounts of $\alpha'$-Ti and $\alpha$-Ti across the three samples, especially between **Figure 6 (a) and (c)**, shows that the amount of $\alpha'$-Ti decreases with temperature, despite the reduced sampling volume available in APT studies.

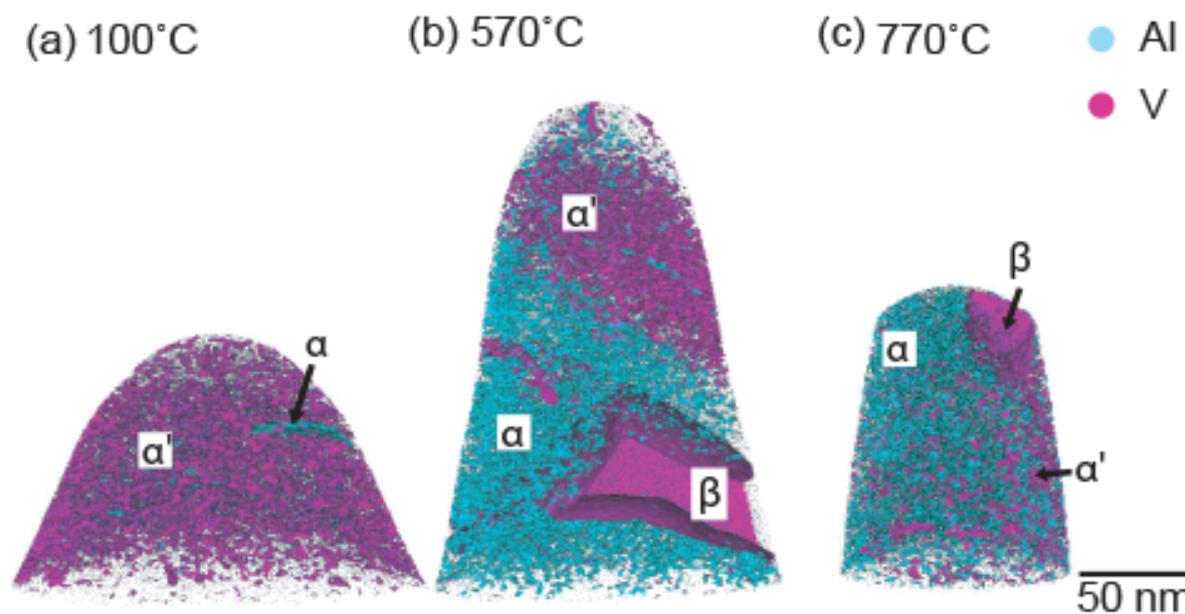

*Figure 6*: *APT reconstruction of the three samples. Isosurfaces of 4 at.% V and 12 at.% Al were applied. The 570°C substrate sample (b) shows the clearest segregation of phases, while the 100°C substrate sample (a) and 770°C substrate sample (c) shows a majority of $\alpha'$-Ti and $\alpha$-Ti microstructure, respectively.*

In order to obtain a reliable breakdown of the elemental composition of each phase, a peak decomposition algorithm was used. This separates the contribution of each constituting element when two peaks overlap by using the natural abundances of isotopes of each element respectively. The composition values obtained for α, β and $\alpha'$ measured on the sample produced by substrate at 570°C, were compared to the literature for similar samples, albeit produced without the heated substrate.

Furthermore, the other vanadium and aluminium-rich regions labelled in **Figure 6 (b)** as α and α′, matched the composition of $α$-Ti and $α′$-Ti found by Tan *et al.* [5] from observation of the 1D concentration profile. The composition of the three phases is tabulated in **Table 1**, while the individual locations of phases are labelled in **Figure 6**. **Table 1** also confirms the absence of $β$-Ti in the sample with a substrate temperature of 100˚C, as the composition of vanadium (β stabilising element) does not exceed 5 at.%.

*Table 1. Compositions (in at.%) of $α′$-Ti, $α$-Ti and $β$-Ti phases obtained from APT data. * denotes the use of a 1D concentration profile instead of decomposition of peaks, while ^ denotes the use of decomposition of peaks but with interfaces created by iso-concentration surfaces.*

|  | Alpha | | | Alpha Prime | | | Beta | |
|---|---|---|---|---|---|---|---|---|
| **Element** | 100ºC* | 570ºC | 770ºC^ | 100ºC* | 570ºC | 770ºC^ | 570ºC | 770ºC |
| Ti | bal. | bal. | bal. | bal. | bal. | bal. | bal. | bal. |
| Al | 11.19 | 10.46 | 14.79 | 6.92 | 9.59 | 10.94 | 2.79 | 1.72 |
| V | 3.21 | 2.37 | 2.30 | 3.85 | 3.69 | 5.13 | 22.69 | 23.79 |
| Cr | 0.13 | 0.28 | 0.20 | 0.14 | 0.77 | 0.91 | 0.72 | 0.16 |
| O | 0.69 | 0.76 | 0.90 | 0.67 | 0.74 | 1.26 | 0.06 | 0.07 |
| Fe | 0.49 | 0.00 | 0.07 | 0.40 | 0.02 | 0.04 | 4.54 | 3.03 |

### 3.4 β phase partitioning at 570˚C

A proximity histogram, in **Figure 7(b)**, applied across the $β$-Ti phase reveals solute partioning with Ti, Al, and O preferentially partitioning to the $α$-Ti phase while V is a $β$-Ti phase stabiliser. This matches Conrad's [20] [21] findings that oxygen and nitrogen are $α$-Ti stabilisers and preferentially partition to the α-phase. A cylinder ROI was applied along a region of high vanadium concentration just next to the large $β$-Ti grain as indicated with an arrow to illustrate the solute segregation at an interface. This region of high concentration is parallel to the grain boundary of the large $β$-Ti grain and could be part of a smaller $β$-Ti grain that is located beyond the sample lifted out. 0.1 at% V and trace (<0.01 at%) concentrations of Fe were detected along linear features assumed to be dislocations in the data set shown in **Figure 7** [17], [22]. At the assumed dislocation, the measured ionic composition of $O^+$ is estimated as 0.016 < 0.020 < 0.024 ionic% (95% confidence interval). There is, however, a known peak overlap of $TiO^+$ and $O^+$ which could allow the underestimate of oxygen content. The peak overlap was analysed further using a custom Matlab program to enable localised (3D) isotopic deconvolution to improve accuracy of atom probe data measurements despite a known local overlap. This analysis showed that there is ~80 times more O from $TiO^+$ than $O^+$ and in this case the error can be ignored [23].

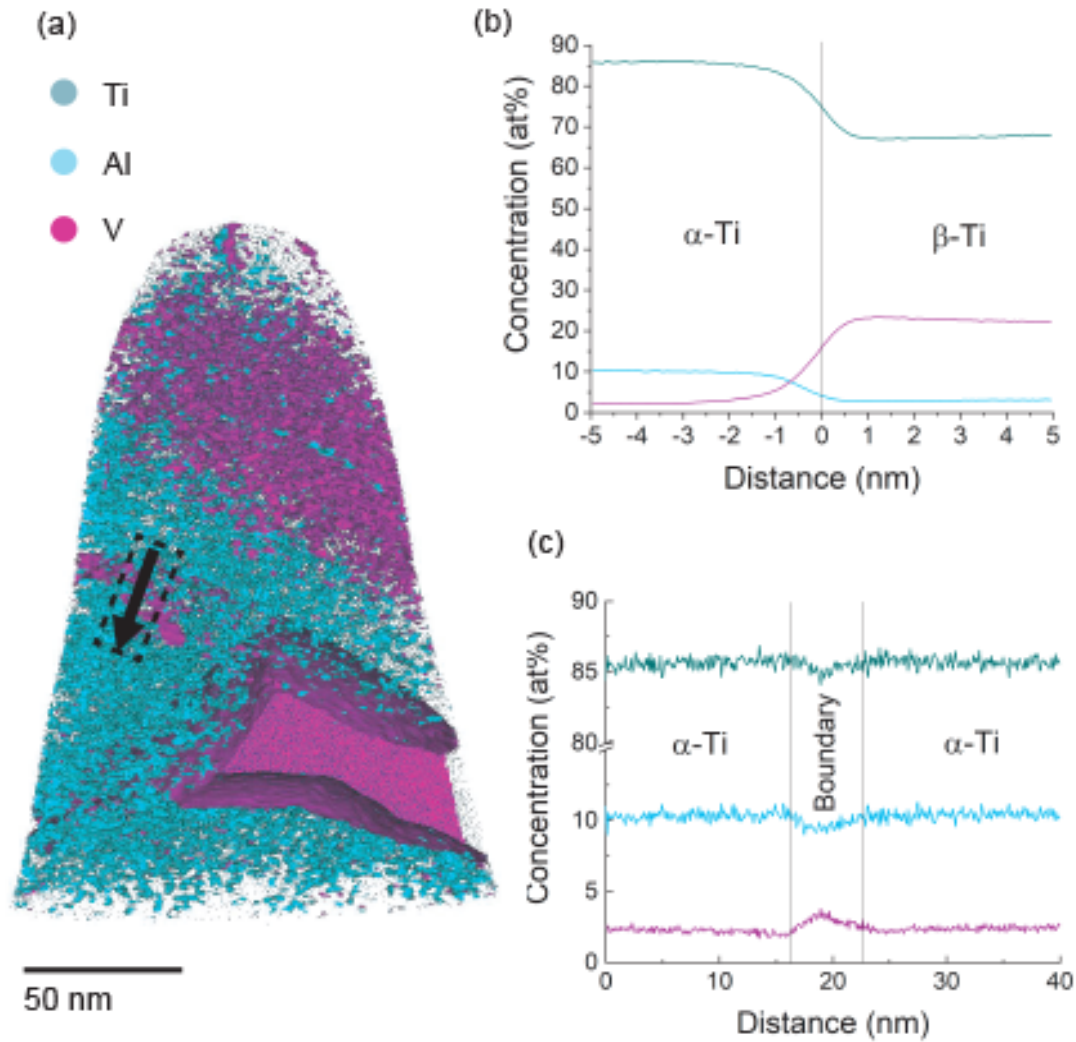

*Figure 7*: APT reconstruction of the 570°C substrate sample (a) with iso-concentration surfaces of 4at% V and 12at% Al applied, and the location of the cylinder region of interest (ROI) demarcated with an arrow. (b) shows the proximity histogram about the large β-Ti grain on the bottom right of the sample, while (c) is a 1D concentration profile of the cylinder ROI indicated in (a).

### 3.5 Segregation of microtwins at 100°C

Iso-density surfaces were applied to the reconstructions of the samples produced at a substrate temperature of 100°C and 770°C, to explore other microstructural features which would not be evident in atom probe reconstructions in the absence of solute segregation, such as dislocations, grain boundaries or twin boundaries, which the TEM micrographs suggested were segregated with Ti and V, as shown in **Figure 3 (c)** and **(e)**. The density iso-surfaces of aluminium and vanadium in **Figure 8 (a)** reveal planar surfaces that appear to only be in specific directions, at ~60° angles from each other, as would be expected of twin boundaries. This suggests that the features follow specific crystallographic planes and together with the micrographs in **Figure 2 (a)**, these can be assumed to be boundaries of twins. The 1D concentration profiles in

**Figure 8 (b)** and **(c)** reveal that there is both titanium and aluminium segregation at each twin boundary.

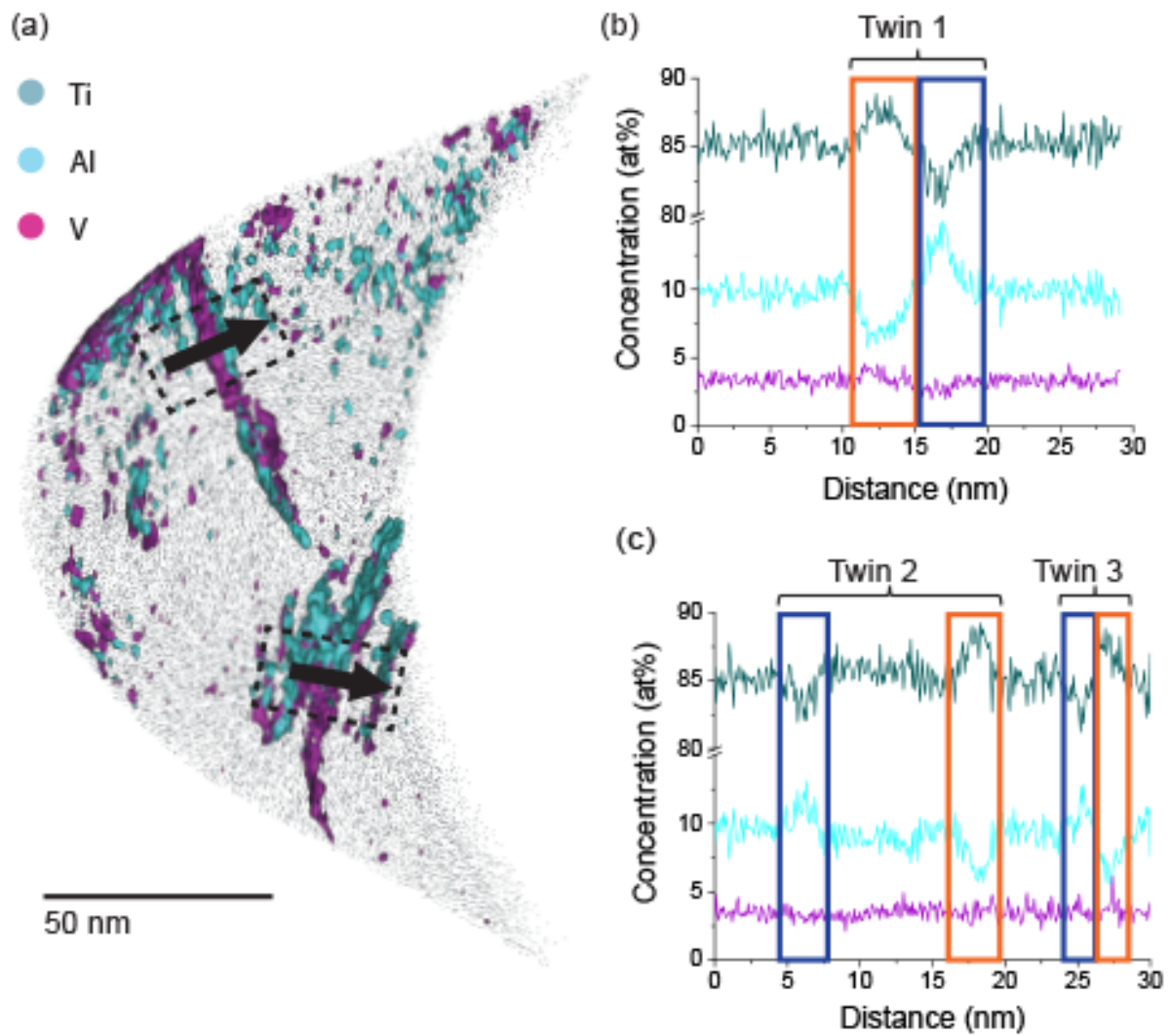

*Figure 8*: (a) Density iso-surfaces of vanadium 2.65/nm³ (magenta) and aluminium 6.90/nm³ (turquoise) were applied to the 100°C substrate sample. Two cylindrical ROIs were added to investigate the planar surfaces, indicated by the dashed line boxes. (b) and (c) show the 1D concentration profile of the two ROIs with their corresponding variation in elemental composition.

### 3.6 Segregation to dislocations at 770°C

In the reconstruction of the sample produced at 770°C, titanium and vanadium isosurfaces delineating regions of high point density were added to reveal the segregation to linear features, which are assumed to be dislocations [17], [22], which could be contributing to pinning and therefore lack of mobility, as observed in the TEM in **Figure 5 (c) to (f)**. A $\beta$-Ti particle can be observed at the top right of the sample, with high-density columnar titanium regions oriented along the z-axis of the sample. The rod-like shape, coupled with the segregation matching the observations performed by TEM, suggests that these are dislocations, while the large quantity of high-density titanium region suggests a significant increase in dislocation density at

770°C. The 1D composition profile in **Figure 9 (c)** helps to quantify the changes in composition detected by the STEM-EDX. It is noted that a similar feature to that of the vanadium rich grain boundary like planar feature is also found next to the β-Ti grain as per the 570°C sample. It is noted that the amount of oxygen in the reconstructions has increased with printing powder bed temperature and could play a role in affecting the mechanical behaviour of the samples.

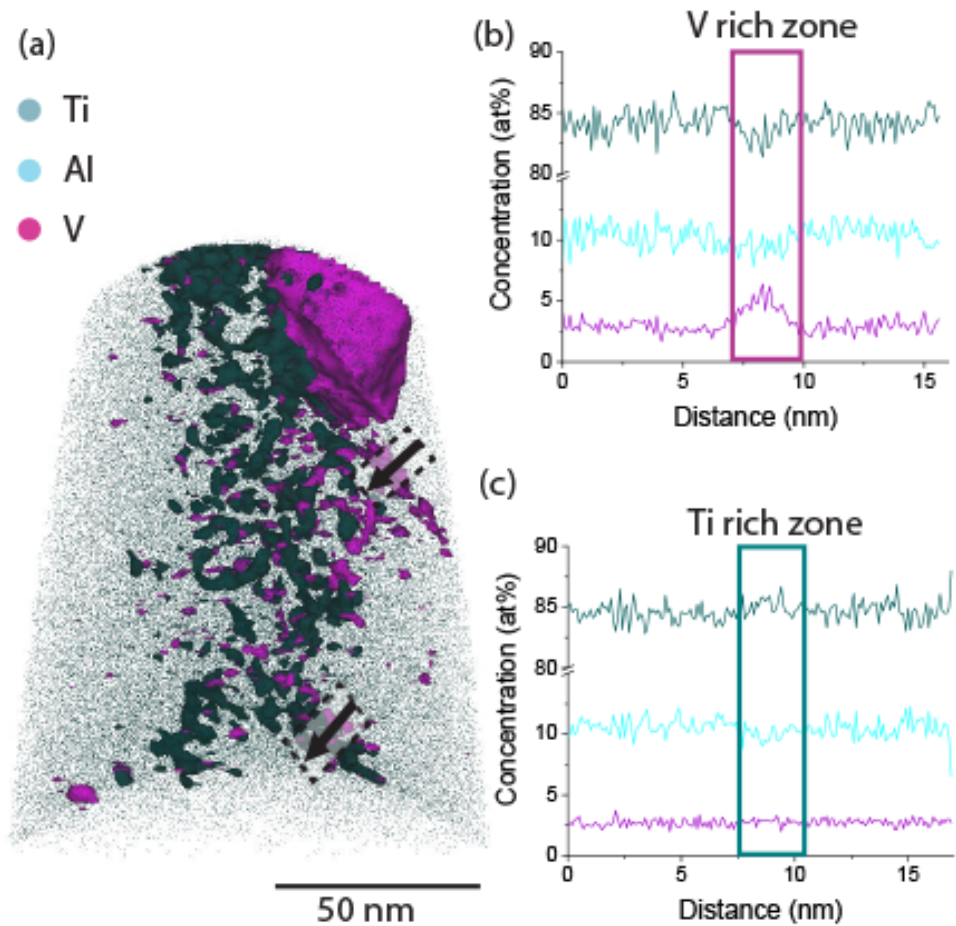

*Figure 9*: *(a) Density isosurfaces of vanadium 2.00/nm³ (magenta) and titanium 51.50/nm³ were applied to the 770°C sample. Two cylindrical ROIs were added to investigate a high-density vanadium region next to the β-Ti grain on the top right and the high-density titanium region, indicated by the dashed line boxes, with their 1D concentration profiles plotted in (b) and (c) respectively.*

## 4. Modelling

To understand the presence of small pockets of β phase in various specimens (e.g. **Figure 5**), solute segregation at dislocations was investigated. In order for localised increases in phase stabilising elements to encourage the formation of new phases, there must be a specific dislocation density needed to provide the driving force for nucleation. In this case, heterogenous nucleation is considered.

The change in the Gibbs free energy of nucleation, $\Delta G$, is described by **Equation 1**:

$$\Delta G = \frac{4}{3}\pi r^3 \Delta G_v S + 4\pi r^2 \gamma_{\alpha\beta} S + \frac{4}{3}\pi r^3 \Delta G_{el} S \qquad (1)$$

Where $r$ is the radius of the nucleus, $\Delta G_v$ is the volumetric driving force, $\Delta G_{el}$ is the elastic strain energy, $\gamma_\alpha$ is the interface energy between α grains and $\gamma_{\alpha\beta}$ is the interface energy between α and β grains.

Nucleation of a new phase would be favourable if $\Delta G$ is negative, as a system always wants to lower its total energy. In order to examine the nucleation model, interface energies and volumetric driving force were determined.

The interface was assumed to be isotropic and incoherent. Gornakova and Prokofjev [24] proposed that interface energies of Ti-64 could be calculated by the equations below:

$$\gamma_\beta = (449 \pm 10) - (0.385 \pm 0.096)(T - T_s) \qquad (2)$$

$$\gamma_\alpha = (2200 \pm 164) - (1.48 \pm 0.20)T \qquad (3)$$

$$\gamma_{\alpha\beta} = (1041 \pm 85) - (0.57 \pm 0.10)T \qquad (4)$$

Where $T$ is the temperature in degrees Celsius and $T_s$ is the solidus temperature, which is 1604 °C for Ti-64. The interface energies at 570 °C (the heated bed temperature in experiments) were thus calculated to be 847.1 mJ/ m² for $\gamma_\beta$, 1362.4 for mJ/ m² for $\gamma_\alpha$ and 716.1 for mJ/ m² for $\gamma_{\alpha\beta}$.

The driving force for the nucleation of the β phase from the α matrix is the difference in Gibbs free energy between the two phases. The Gibbs free energy of each phase of Ti-64 can be calculated by:

$$G^\beta = G^0 + G^{ideal} + G^{XS} \qquad (5)$$

Where:

$$G^0 = X_{Al} G0_{Al}^{BCC} + X_{Ti} G0_{Ti}^{BCC} + X_V G0_V^{BCC} \qquad (6)$$

$$G^{ideal} = RT(X_{Al} \ln X_{Al} + X_{Ti} \ln X_{Ti} + X_V \ln X_V) \qquad (7)$$

$$\begin{aligned} G^{XS} = & X_{Al}X_{Ti}(-125485 + 36.8394T) + X_{Al}X_V[(-95000 + 20T) + (-6000)(X_{Al} - X_V)] \\ & + X_{Ti}X_V[(10500 - 1.5T) + 2000(X_{Ti} - X_V) + 1000(X_{Ti} - X_V)^2] \\ & + X_{Al}X_{Ti}X_V[X_{Al}(116976.3 - 9.067T) + X_{Ti}(-175169 + 59T) \\ & + X_V(31107.3 - 42.316T) \end{aligned}$$

$$(8)$$

Here, $G^0$ is the Gibbs free energy of a mechanical mixture of the constituents, $G^{ideal}$ is the entropy of mixing for an ideal solution, $G^{XS}$ is the excess term and X is the mole fraction of each element in Ti-64 [25]. The mole fraction of V, Ti and Al at the dislocation core was estimated to be 0.04, 0.86 and 0.1 respectively. Other parameters in the equations could be found from SGTE database [26].

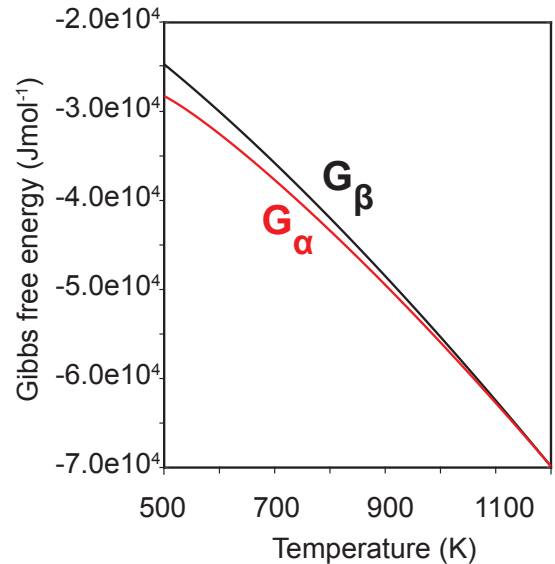

*Figure 10: Variation of Gibbs free energy with temperature, where $G_α$ represents the Gibbs free energy of the α phase and $G_β$ represents the Gibbs free energy of the β phase.*

From **Figure 10**, it is observed that below the β transus (980 ˚C/1253 K), the Gibbs free energy of the α phase is always lower than that of β phase. Thus, the α phase is more stable and nucleation of β phase is unfavourable. However, Suprobo et al. [27] proposed that the presence of dislocations could help raise the Gibbs free energy of α phase and the contribution can be calculated by:

$$\Delta G_s = \varepsilon \rho V_m \tag{9}$$

where ε is the strain energy and $V_m$ is the molar volume of Ti-64 ($1.06 \times 10^{-5}$ m³/mol). The strain energy is calculated by:

$$\varepsilon = \frac{\mu b^2}{4\pi} \ln \frac{R_e}{r} \tag{10}$$

where $\mu$ is the shear modulus of Ti-64 ($4.4 \times 10^{10}$ N/m²), $r$ is the radius of dislocation core which is assumed to be 1 $b$, where $b$ is the Burgers vector, and $R_e$ is the outer cut-off radius of dislocation core ($=\frac{1}{\sqrt{\pi\rho}}$).

To understand how the addition of dislocations in the matrix affects the Gibbs free energy, and the energy for nucleation of a new phase, the dislocation velocity and solute diffusion rate of V were calculated, as V was observed segregating to defects at 570 °C.

Dislocation velocity can be calculated by:

$$v = \frac{\varepsilon \ (strain\ rate)}{\rho b} \tag{11}$$

where ρ is the dislocation density and $b$ is the Burgers vector [28]. The strain rate was assumed to be dependent of the cooling rate of 30°Cmin$^{-1}$, and was estimated to be 1.4 x 10$^{-3}$ s$^{-1}$ [29]. The dislocations were assumed to glide in the basal plane and the Burgers vector was found to be 0.24×10$^{-9}$ m. The dislocation density was estimated using the 'line-intercept method' to be 1.09×10$^{14}$ m$^{-2}$ [30]. Thus, the dislocation velocity was calculated to be 4.23×10$^{-4}$ m/s at 570 °C.

The diffusion of V in α matrix was assumed to be in a direction perpendicular to c-axis of the HCP lattice. The diffusion coefficient of steady state diffusion is normally represented by the Arrhenius equation:

$$D = D_0 \times e^{-\frac{Q}{RT}} \tag{12}$$

where $D_0$ is the pre-factor, Q is the activation energy, R is the gas constant and T is the temperature. For diffusion of V in α-Ti at 570 °C, $D_0$ was found to be 2.33×10$^{-18}$ m$^2$/s and Q was found to be 12970.4 J [31]. However, Ti-64 is a ternary system, and Lindwall et al. [32] suggested that the presence of Al would affect the diffusion rate of V and the effect was shown by $D_2 = 10^{-0.12Al + log\ D}$. The mole fraction of Al in Ti-64 was estimated to be 10 % and thus the experimental diffusion coefficient of V was calculated to be 2.31×10$^{-20}$ m$^2$/s.

When the dislocation motion is controlled by the drag of solute atmosphere, which means that the solute cloud diffuses along with the dislocation and segregation occurs as shown by **Figure 5**, the relationship between the dislocation velocity and solute diffusion coefficient can be summarized by:

$$v = v_d \frac{\tau b^4}{kT} \exp(-\frac{U_d}{kT}) = D \frac{\tau b^2}{kT} \tag{13}$$

where τ is the applied stress which is normally in the order of 10$^9$ [33]. Applying a dislocation velocity of 4.23×10$^{-4}$ m/s, for solute segregation to occur the theoretical minimum diffusion coefficient of V was found to be 8.63×10$^{-14}$ m$^2$/s. The experimental diffusion coefficient (2.31×10$^{-20}$ m$^2$/s) is slower than the theoretical minimum.

By taking into account the contribution of dislocation strain, a new Gibbs free energy curve for the α phase was plotted as shown by the dashed blue curves in **Figure 14**.

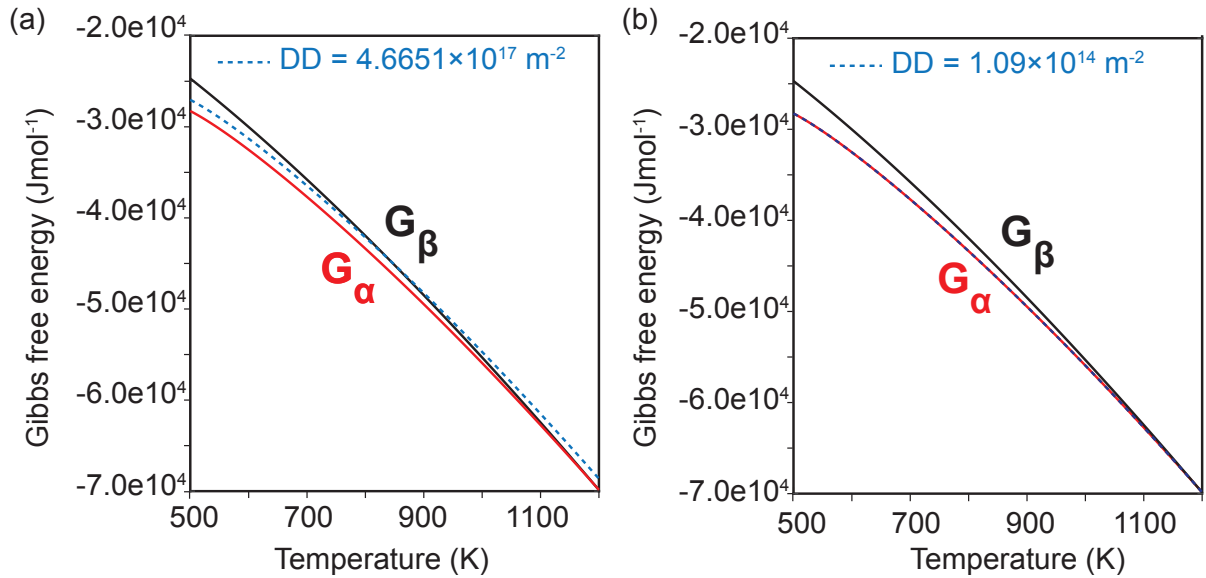

***Figure 14***: *Variation of Gibbs free energy with temperature when dislocation density is a) 4.6651×10$^{17}$ /m$^2$ and b) 1.09×10$^{14}$ /m$^2$.*

At 570 °C, the Gibbs free energy of β phase is greater than that of α phase by 1.2253×10$^3$ J/mol. In order to trigger the nucleation of β phase, the dislocation strain needs to raise the free energy of α phase by that amount. Using **Equation 14**, the minimum dislocation density required was calculated to be 4.6651×10$^{17}$ /m$^2$. However, the dislocation density found experimentally in our sample was only 1.09×10$^{14}$ /m$^2$, far smaller than the theoretical value. Thus, the Gibbs free energy of α phase is still lower as shown in **Figure 14 (b)** and nucleation of β phase is unlikely as negative driving force could not be established. However, nano-scale $β$ precipitates were observed in the TEM micrographs. This is probably due to the ununiform dislocation distribution throughout the sample. Some parts of the sample have dislocation density high enough to trigger the nucleation of β phase.

## 5. Discussion

### 5.1 Effect of substrate temperature on microstructure evolution

One of the most interesting results from this work is the significant and somewhat surprising- differences in the microstructure when varying the substrate temperatures. Table 3 summarises such changes. At 100°C, very little β forms and there is a large phase fraction of α'. Additionally, there are significantly more pores per unit area, with the highest volume fraction of pores of all the tested samples. The microstructure contains both α and α', and has a lower dislocation density than the other two conditions. The rapid cooling rate upon solidification provides necessary driving force for large scale β to α' transformation [34], [35] and shows a fast cooling microstructure with the presence of microtwins [36]. The very low fraction of β phase is likely due to rapid solidification and little diffusion at 100°C preventing its formation. In addition, there is a presence of microtwins with both Al and V rich regions, though there is not both Al and V rich regions together at the interface. This indicates that short range diffusion is occurring due to the formation of these microtwins within the alloy, which could be due to a partitioning effect towards forming α and β in the Al- and V-rich regions; this mechanism has been seen experimentally in other alloy systems [37],

[38]. Microtwins are a result of the high lattice strains and low symmetry between the α and β phases [39], therefore chemical segregation occurs.

*Table* . *Predicted local Al, V and O concentrations at crystal defects and their influence on local variations in solid solution strengthening.*

|  | 100 °C | 570 °C | 770 °C |
|---|---|---|---|
| Primary phases | α'+α | α'+α+(nano)β | α+β |
| Crystal defects | Twinning | Dislocation networks | Dislocation tangles |
| Solute behaviour | Al, V segregation at twins | V segregation at dislocations | V segregation at dislocations |

Upon increasing the substrate temperature to 570°C, there is no longer evidence of microtwinning. Instead, nano-scale β are observed within α' and α. Though the largest pore radius is seen at this temperature, the ductility is the highest of all 3 samples discussed here. Dislocation networks were observed within the α lamellae, which may be due to the absence of microtwinning and the presence of β at the α-grain boundaries. Interfacial dislocations in Ti-64 may form at the (moving) α'/β interface as α' nucleates and grows during rapid solidification in the first build cycle, due to the high shear strain of the martensitic transformation and the high local lattice mismatch between α', α and β phases. The substrate temperature is high enough to promote dislocation recovery, however the remnant β at the grain-boundary should inhibit significant dislocation annihilation and local dislocation reconfiguration could take place instead. As for how the networks influence mechanical properties, they could act as "soft barriers" for slip [38]. Using the analogy of dislocation networks forming in single-crystal (SX) Ni-based superalloys, Rai *et al.* have observed that the formation of dislocation networks at the $\gamma/\gamma'$ interfaces promotes strain localisation and material softening under low-cycle fatigue [40]; the deformation temperature was 850°C, which is sufficiently low to prevent directional coarsening and dislocation reconfiguration was primarily driven by the interfacial misfit and deformation within the matrix. This indicates that in the present case, the dislocation networks promote an increase in ductility by allowing more plasticity to take place locally within the α lamellae, whereas the fine grain boundaries and nano-β should compensate for the local softening and keep the macroscopic strength high. In addition, the micro texture of the alloy from the EBSD shown in **Figure 5** indicates that there may be <a> type dislocations which could be transferred easily between α lamellae, which may be further increasing the ductility of the material, however this is similar in all samples and is often found in similar titanium microstructures [41], [42].

At 770°C, there is a significant change in mechanical properties and this is reflected also in the dislocation content and phase structure. A rapid drop in ductility is observed, despite the specimen containing the smallest pores. There is substantially more β within this sample and very little α' remains, however, this should not lead to such a large loss in ductility. Dislocation tangles are observed in **Figure 8**; these could be formed by a similar mechanism as the dislocation networks at 570°C but the higher build temperatures could promote extensive recovery and partial loss of the networks; the high density of dislocation networks observed in the vicinity of the α/β interface in **Figure 8** supports this mechanisms, as grain-boundary β should limit the extent of dislocation recovery by maintaining the interfacial misfit. The lack of ductility in this sample indicates that the dislocation structures are no longer acting as soft barriers for dislocations and additional mechanisms are operating which could hinder slip. The fact that V-rich dislocations are observed (**Figure 12**) indicates that solute redistribution is playing an important role in controlling dislocation slip, however it could also be stated that O has a role to play in leading to such a brittle sample [43].

The presence of H and/or O due to process-related contamination could affect strongly the elongation at 770°C, as these elements are well known to promote substantial reductions in ductility in titanium alloys [20]. It is common for samples produced via LBPF to contain increased oxygen content, particularly within the α' phase [44]. This can be explained with the higher than required partial pressure of oxygen in the LBPF machine which leads to oxygen uptake during printing []. For instance, Tan *et al*. [5] measured high H and O levels (between 0.1-1 at%) in Ti-6Al-4V produced by EPBF. Oxygen content was quantified through atom probe tomography (APT). APT was used to identify microstructural features according to their chemical composition and quantify the elemental segregation at interfaces and dislocations observed in the TEM in **Figure 6 to 8**. It has also been previously indicated that oxygen increases the critical stress for martensite formation, and increased oxygen content may suppress martensite formation. As increased substrate temperatures have less martensite, this could indicate that heating the substrate results in higher oxygen contents of the alloy [45]. The presence of any number of elements may aid the atomic shuffle required for the martensitic transformation, and therefore local composition is highly important.

## 5.2 Comparing against microstructures obtained by other Additive Manufacturing methods

To understand better how the preheating temperatures affect local changes in thermal history and microstructure, it is worth comparing the results from this work with results obtained by processing Ti-6Al-4V using other AM technologies. At 100 °C, the microstructure is comparable to as-built microstructures obtained by conventional LBPF methods in as-built conditions, which have reported very high cooling rate ($\sim 10^6$ K/s) and lowest build temperature [46]. Yang et al. observed a high density of dislocation tangles, twins and fine-scale β; they argued that fine-scale β is unstable due to their XRD results did not show β peaks [47]. The authors did however observe a shift in (110) peaks compared to calculated lattice parameters, which they attributed to the increased solubility of V in α', which was not confirmed experimentally; other authors have argued that the peak shift is caused by solute redistribution [10]. Similar microstructures have been reported by several authors [46], however no elemental segregation at twins has been reported before.

On the other hand, the observed microstructure at 570°C is partially similar to microstructures obtained by LBPF after heat treating the samples at medium to low temperatures. For instance, Xu *et al*. [8] used SEM to characterise the microstructure of SLM Ti-6Al-4V in as-built state and after applying different heat treatments. The Backscatter Electron images (BSE) of the as-built samples did not show evidence of forming nano-scale β at grain boundaries, however when post-LPBF tempering at 540 °C, isolated nano particles of β along the α boundaries were observed. The authors also reported that the decomposition of martensite takes place at temperatures as low as 400 °C. No elemental segregation at dislocations was reported.

The observed microstructure at 770°C is similar to microstructures obtained by Electron Beam Melting (EBM), where the cooling rate is less severe (~$10^4$ K/s) and the local build temperature is comparable [48]. For instance, Tan *et al.* [5] studied the transition of β to α' and α during EBM by means of Atom Probe and SEM. To do this they varied the thickness of the built samples to determine the variations in microstructure, thin samples where fully martensitic and thick samples contained primarily α+β. Their Atom Probe results showed local elemental enrichment (V) at dislocations in martensitic samples (thin) but no segregation at twins was reported, whereas the composition in the thick samples corresponded to that at equilibrium in an α+β microstructure. They argued that discrete β particles form initially along the α' plate boundaries and grow subsequently, however they only referred to the low possibility of finding connecting β rods within the α' plates.

In summary, the microstructures of titanium alloys are incredibly sensitive to thermal processing and thermal history. Parameters such as cooling rate, both above and below the β transus, are critical to the resultant microstructure, which is directly linked to the final strength and fatigue performance [43], [49]. Extended time at temperature due to substrate heating as seen here will therefore act as pseudo ageing, comparable to standard processes. This has a significant effect on the local chemistry and diffusion and how this interacts with defects within the material, as well as microstructural evolution.

**5.3 Phase kinetics and relation with martensite transformation parameters.**

The phase transformation sequence in Ti-6Al-4V produced by LBPF and EBM differs during the final stages but is similar during the initial stages of the thermal cycle [5], [47], which can serve as baseline for analysing the transformation sequence in the present work. The sequence for LBPF and EBM is as follows: during the first cycle β forms upon solidification which rapidly transforms into α' due to the very high cooling rates; during subsequent cycles, α' partially transforms back to β upon reheating, which then transforms again to α' upon cooling, leading to primary and secondary α'; a small fraction of retained β remains during the build process and α' eventually transforms to α+β in EBM. In order to understand the differences in the primary phases observed at different preheating temperatures (**Table 3**), it is important to also identify the transformation kinetics of β to α and β to α' in Ti-6Al-4V. The martensite start, $M_s$, and finish, $M_f$, temperatures have been estimated as ~790°C and ~710°C, respectively [50], whereas the α start , i.e. β transus, and finish  are 980°C and ~570°C, respectively.

At 770°C, the substrate is just below the $M_s$ and above $M_f$, therefore much less α' forms during the first and subsequent thermal cycles and the substrate spends a significantly longer time in the α + β phase field, leading to more energy being supplied for diffusion. As α is a more stable a preferable phase that α', and more prevalent at slower cooling rates, a lack of α' is observed in this specimen. Possible slower cooling additionally means that as the material is cooled through the β transus, mass phase change occurs. This is evident is the tangled dislocations in **Figure 8** and the small β particles that are observed. Geometrically necessary dislocations are prevalent in these types of microstructures due to the misfit between the α and β phases [17], [51], where the β phase is much larger, resulting in both a strain and rotation to create a favourable low energy interface. Once the β phase has transformed upon cooling, these dislocations will still be present, and furthermore may explain the partitioning of β stabilisers to dislocations.

At 570˚C, higher undercooling promotes significant α' formation and, since the preheating is at , α nucleation will also be thermodynamically favourable but after longer times; as the preheating temperatures is below $M_f$, α will likely nucleate from the α' by solute partitioning. However, at this temperature the energy for chemical diffusion is lower, providing a much slower growth velocity for the α phase [16], [52]. Nano β is observed at this substrate temperature and is less prevalent that at 770˚C, however APT shows an increase of vanadium at the α/α boundary which could indicate a much finer presence of β within the microstructure.

This analysis also indicates that for preheating temperatures below 570˚C, e.g. at 100˚C, the microstructures will be primarily martensitic with α eventually forming and a small fraction of β might be possible if there is enough thermal energy for elemental diffusion; α' will decompose above , as α nucleation will be more prolific, therefore leading primarily to α+β. Our analysis in **Table 3** is consistent with these conclusions.

As for other microstructural features, microtwinning in α' is only observed when preheating at 100˚C, whereas the IPF maps for 570˚C and 770˚C in **Figure 5** showed a somewhat resemblance of an initially twinned microstructure but it has evolved into lamellar α+β. This indicates that in-situ recovery has taken place at higher preheating temperatures, however it was not possible to find information on the recovery kinetics of twinning in Ti-6Al-4V during high temperature annealing. Nonetheless, if the athermal transformation from β to α' is rationalised as an introduction of shear deformation in the sample, we can qualitatively compare the behaviour of twinning observed in the present work with that for mechanical twinning in Ti-6Al-4V deformed at different temperatures. Mechanical twinning in Ti-6Al-4V typically occurs only when it is deformed at very high strain rates or very low temperatures [53]. Under such conditions, mechanical twinning has been reported to form up to a maximum deformation temperature of ~400˚C-450˚C. Hao et al [54] reported in the near-α alloy Ti-6Al-2Zr-1Mo-V wt% that twinning forms under tension at and below 400˚C. Similarly, Zeng et al [55] reported in commercially pure Ti that twinning is suppressed at 650˚C and Fitzner et al [56] reported that twinning forms at room temperature in near-α Ti-Al alloys with Al up to 12at%, whereas for higher Al contents it is suppressed due to the activation of basal slip. Therefore, since twinning impedes dislocation motion and potentially lowering material's ductility, if the preheating conditions are above 400˚C-450˚C the ductility can in principle increase via promoting twin recovery.

### 5.4 Effects of solute segregation at defects

### 6. Conclusions

Samples of Ti-6Al-4V were produced through LPBF on a heated substrate between 100-770°C. Increasing the temperature of the substrate from 100°C to 570°C improved the ductility during room temperature tensile tests, with 570°C being the ductility maximum. Further heating of the substrate during sample production up to 770°C reduced ductility to zero. Although the ductility changed significantly with processing conditions, the UTS at 100°C and 570°C was approximately constant (~1.2GPa) and it had the lowest value at 770°C, probably due the alloy's brittle behaviour. Our analysis employed SEM, TEM and APT to study the complex variations in microstructure and solute behaviour to establish the specific mechanisms controlling the strength and ductility at different substrate temperatures. In summary:

- At 100°C, a heavily strained and twinned microstructure, primarily composed of α+α', was observed and it was comparable to as-built microstructures obtained by conventional LPBF methods. The low ductility this sample was attributed to the high density of microtwins and dislocations preventing any further plasticity and/or dislocation slip. At 570°C, twins are no longer present and instead nano-scale β precipitates are observed within α' and α, as well as dislocation networks. The lack of twins was attributed to this temperature being higher than the temperature for twin recovery in Ti-6Al-4V (~400°C-450°C). The microstructure at 570°C was partially similar to microstructures obtained by conventional LPBF after heat treating the samples at medium to low temperatures. At 770°C, there is substantially more β, very little α' remains, and dislocation tangles form within the α grain interiors. The observed microstructure at 770°C is similar to microstructures obtained by Electron Beam Melting, where the local build temperature is comparable and the cooling rates are less severe, leading to a mix of α+β and small residual α'.
- The dislocation networks observed in the 570°C sample may have formed by a sequence of interfacial dislocations generating at the α'/β interface as α' nucleates during rapid cooling. The substrate temperatures were high enough to promote dislocation recovery but grain-boundary β precipitates could have inhibited significant dislocation annihilation and local dislocation reconfiguration takes place instead. It was concluded that dislocation networks in the 570°C sample act as "soft barriers" for slip and help in increasing the ductility.
- The dislocation tangles observed at 770°C could have formed by a similar mechanism but the higher build temperature promotes extensive recovery and partial network dissolution. Evidence of this was that well-defined networks were still present in the vicinity of the α/β interfaces, where grain-boundary β should limit the extent of dislocation recovery by maintaining the interfacial misfit. The lack of ductility at 770°C was attributed to local solute redistribution causing dislocation pinning and an increase of O content in this sample, particularly at α' where a high dislocation density is present.
- Solute segregation at crystal defects was observed in all pre-heating conditions. Al and V segregation at microtwins was observed in the 100°C sample, with mutually exclusive Al- and V-rich regions forming in adjacent twins. This indicates that the shear strain induced by twinning induces a driving force high enough for solute partitioning and short range diffusion towards forming α and β in the Al- and V-rich regions. To the authors knowledge, it is the first time such "selective" solute partitioning at twins is observed in Ti.
- V segregation at dislocations was observed in the 570°C and 770°C samples, consistent with the higher preheating temperatures. The observations are in agreement with previous reports in Ti-6Al-4V produced by Electron Beam Powder Fusion but, to the authors knowledge, this is the first work reporting solute segregation at dislocations in Ti-6Al-4V using Laser-based technologies.

- High O contents were measured in all samples. At 570°C and 770°C, Oxygen promotes an increase in the solid solution strengthening of ~375-300 MPa, potentially aiding in maintaining the strength high. This can be reflected in the macroscopic strength being nearly identical at 570°C and 100° C, in spite of the microstructure being coarser in the former. However, O contents at 770°C were just above the threshold for O embrittlement, therefore the sample showed a lack of ductility. At 100°C and 570°C, the O was measured to be below the critical threshold.
- Based on the complicated phase transformation sequences and microstructural variations analysed this work, optimal in-situ heat treatments for improved microstructural control and mechanical properties should likely be in the temperature range of 450°C-570°C. This processing window results in a strong (~1.2GPa) and fine α'+α+β structure containing a high density of dislocation newtorks, acting as soft barriers for slip increasing the ductility. The relatively-high temperatures also promote short-range diffusion for solute segregation and localised solid solution strengthening for improved strength. This temperature range should produce samples below the critical Oxygen content for severe embrittlement, whilst the O adsorbed during the build process may promote additional solid solution strengthening, although the actual levels of O absortion are material, machine and process-specific.

**Acknowledgements**
E.I. Galindo-Nava, S. Pedrazzini and T. B. Britton would all like to acknowledge the Royal Academy of Engineering for their research fellowships. E.I. Galindo-Nava performed this work under EPSRC grant EP/T008687/1. S. Pedrazzini performed part of this work under EPSRC grant EP/M005607/1 and under EPSRC fellowship EP/S0138881/1. Giorgio Divitini is gratefully acknowledged for his help with TEM. P. Bajaj acknowledges funding by the DFG under grant number JA2482/2-1. EPSRC Future Manufacturing Hub in Manufacture using Advanced Powder Processes (MAPP)(EP/P006566/1) is acknowledged for their support during this investigation. T. Dessolier and T. B. Britton acknowledge funding from the Shell-Imperial Advances Interfacial Materials Science (AIMS) Centre.